\documentclass[10pt,a4paper,twoside]{article}
\usepackage{epsfig}
\usepackage{baltlat6}
\usepackage{array}
\usepackage{here}
\usepackage{graphics}
\usepackage{graphicx}
\usepackage{epstopdf}

\begin{document}
\ \
\vspace{0.5mm}
\setcounter{page}{100}

\titlehead{Baltic Astronomy, vol.\,24, 100--108, 2015}

\titleb{THE FIRST LIGHT OF Mini-MegaTORTORA WIDE-FIELD\\ MONITORING
SYSTEM}

\vskip-2mm

\begin{authorl}
\authorb{A. Biryukov}{1,4},
\authorb{G. Beskin}{2,4},
\authorb{S. Karpov}{2,4},
\authorb{S. Bondar}{3},
\authorb{E. Ivanov}{3},
\authorb{E. Katkova}{3},
\authorb{A. Perkov}{3,4} and
\authorb{V. Sasyuk}{4}
\end{authorl}

\begin{addressl}
 \addressb{1}{Sternberg Astronomical Institute of M.\,V. Lomonosov
Moscow State University, 13 Universitetskij pr., Moscow, 119991 Russia}
\addressb{2}{Special Astrophysical Observatory, Karachai-Cherkessia, Nizhnij Arkhyz,\\ 369167 Russia}
\addressb{3}{Precision Systems and Instruments Corp., 53 Aviamotornaya
str., Moscow,\\ 111024 Russia}
\addressb{4}{Kazan Federal University, 18 Kremlevskaya str., Kazan, 420008 Russia}
\end{addressl}

\vskip-2mm

\submitb{Received: 2014 December 20; accepted: 2015 January 6}

\vskip-2mm

\begin{summary} We describe the first light of a new 9-channel
wide-field optical monitoring system with sub-second temporal
resolution, Mini-MegaTORTORA, which is being tested now at the Special
Astrophysical Observatory in Russian Caucasus.  The system is able to
observe the sky simultaneously in either wide ($\sim$\,900 deg$^2$) or
narrow ($\sim$\,100 deg$^2$) fields of view, either in clear light or
with any combination of color (Johnson $B$, $V$ or $R$) and polarimetric
filters installed, with exposure times ranging from 100 ms to 100 s. The
primary goal of the system is the detection of rapid (with sub-second
characteristic time scales) optical transients, but it may be also used
for studying variability of sky objects over longer time scales.
\end{summary}


\resthead{The first light of Mini-MegaTORTORA} {A. Biryukov, G.
Beskin, S. Karpov et al.}

\sectionb{1}{INTRODUCTION}

Mini-MegaTORTORA is a new robotic instrument just commissioned for the
Kazan Federal University and developed according to the principles of
MegaTORTORA multi-channel and transforming design formulated by us
earlier (see Beskin et al. 2010a; Karpov et al. 2012, and references
therein).  It is a successor to the FAVOR (Zolotukhin et al. 2004;
Karpov et al. 2005) and TORTORA (Molinari et al. 2006) single-objective
monitoring instruments we built earlier to detect and characterize fast
optical transients of various origins:  cosmological, galactic, and
near-Earth.  The importance of such instruments became evident after the
discovery and detailed study of the ever-brightest optical afterglow of
a gamma-ray burst, GRB080319B (Karpov et al. 2008; Racusin et al. 2008;
Beskin et al. 2010b).

The Mini-MegaTORTORA (MMT) system includes a set of nine individual
channels (see Fig.~1) installed in pairs on equatorial mounts (Figs. 2
and 3).  Every channel has a coelostat mirror installed before the Canon
EF85/1.2 objective for a rapid (faster than 1~s) adjusting of the
objective direction in a limited range (approximately $10^\circ$ to any
direction).  This permits either a mosaic of the larger field of view or
pointing all the channels in one direction.  In the latter mode, a set
of color (Johnson $B$, $V$ or $R$) and polarimetric (three
different directions) filters can be inserted before the objective to
maximize the information acquired for the observed region of the sky
(performing both three-color photometry and polarimetry).


\begin{figure}[!t]
  \vbox{
    {\centering
\resizebox*{\columnwidth}{!}{\includegraphics[angle=0]{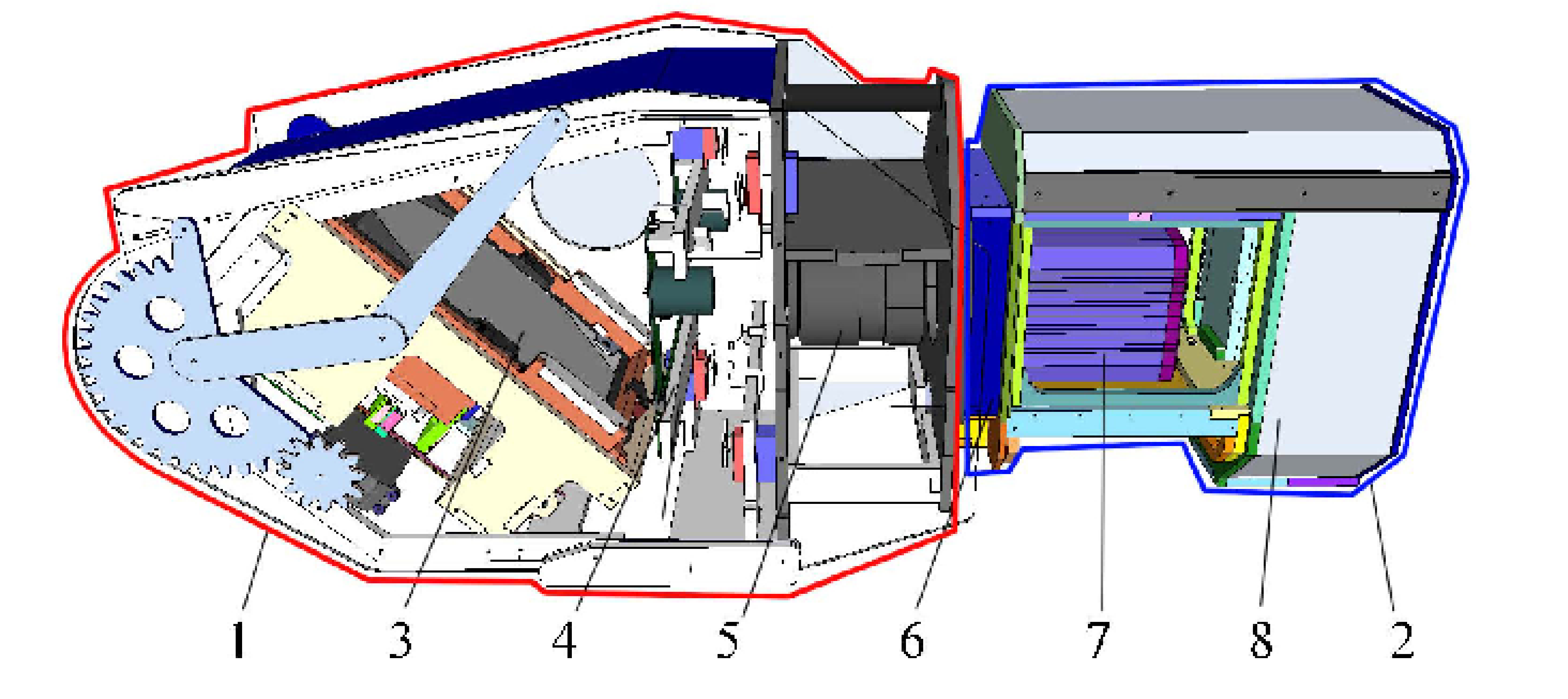}}
    }
    \vspace{-1mm}
\captionb{1} {Schematic view of a single MMT channel.  {\it 1}\,:
Coelostat unit; {\it 2}\,:  camera unit; {\it 3}\,:  coelostat mirror
that can rotate by $\sim10^\circ$ around two axes; {\it 4}\,:
installable color and polarimetric filters; {\it 5}\,:  Canon EF85/1.2
objective; {\it 6}\,:  optical corrector; {\it 7}\,:  Andor Neo sCMOS
detector; {\it 8}\,:  conditioner keeping stable environmental
conditions inside the channel.} }
\end{figure}
\vskip5mm
\begin{figure}[!tH]
  \vbox{
   {\centering
\resizebox*{0.59\columnwidth}{!}{\includegraphics[angle=0]{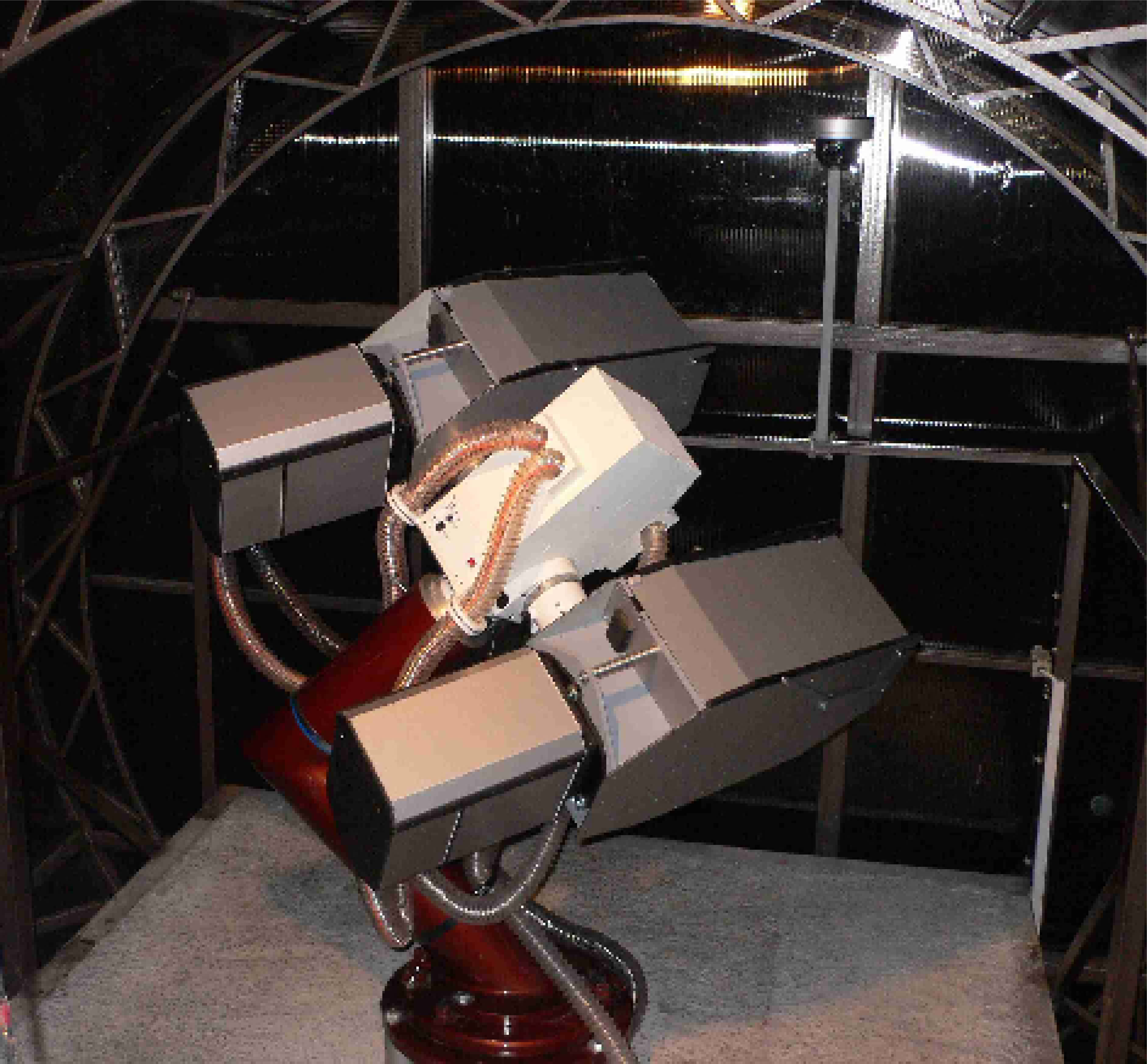}}
\resizebox*{0.41\columnwidth}{!}{\includegraphics[angle=0]{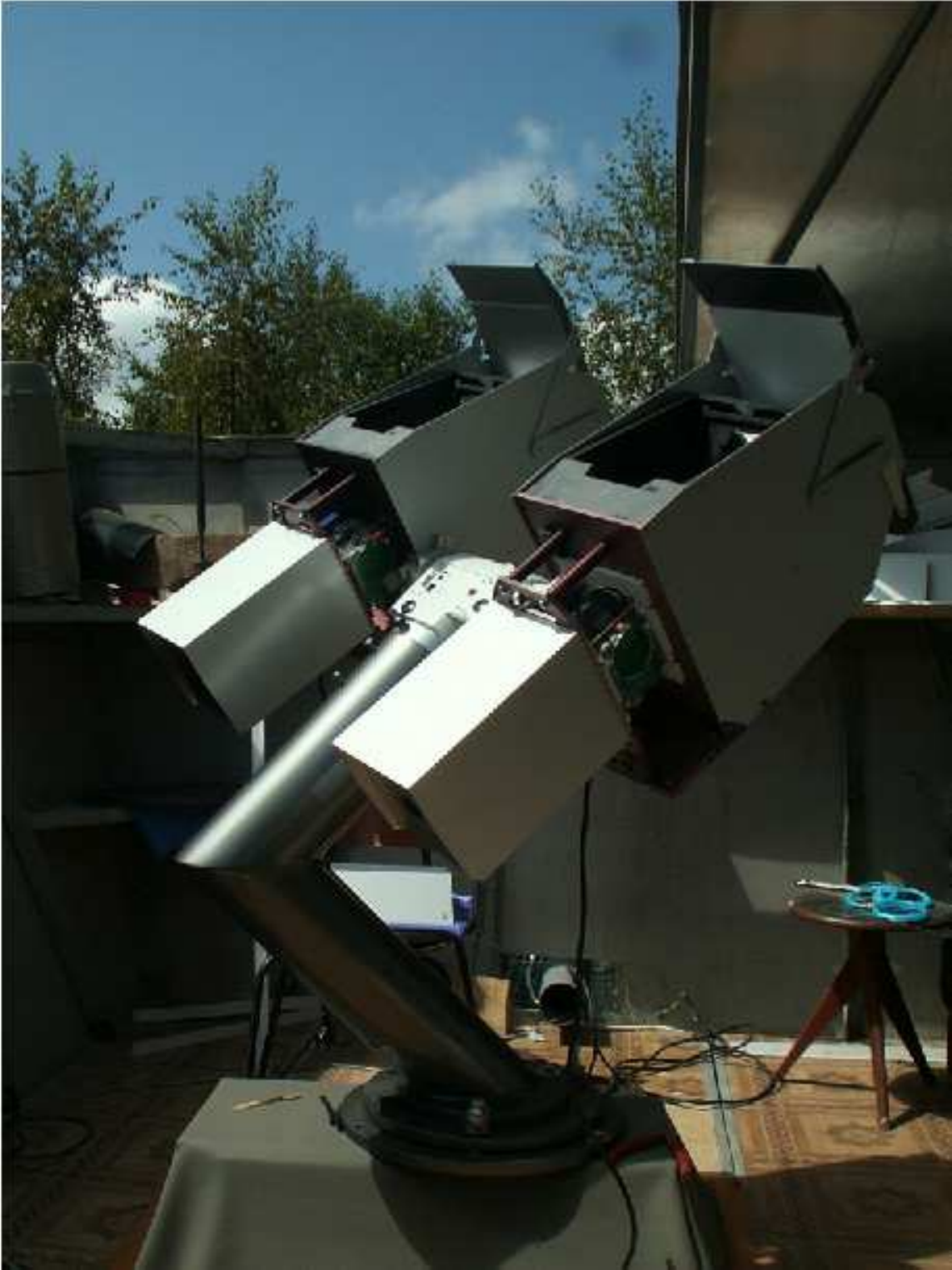}}
   }
\vskip2mm
\captionb{2} {Two MMT channels installed on a single mount.  The
complete system consists of five such mounts, carrying nine operative
channels and one replacement channel.}
}
\vspace{6mm}
\end{figure}


\begin{figure}[!tH]
\vbox{
	{\centering
    \begin{center}
      \resizebox*{\columnwidth}{!}{\includegraphics[angle=0]{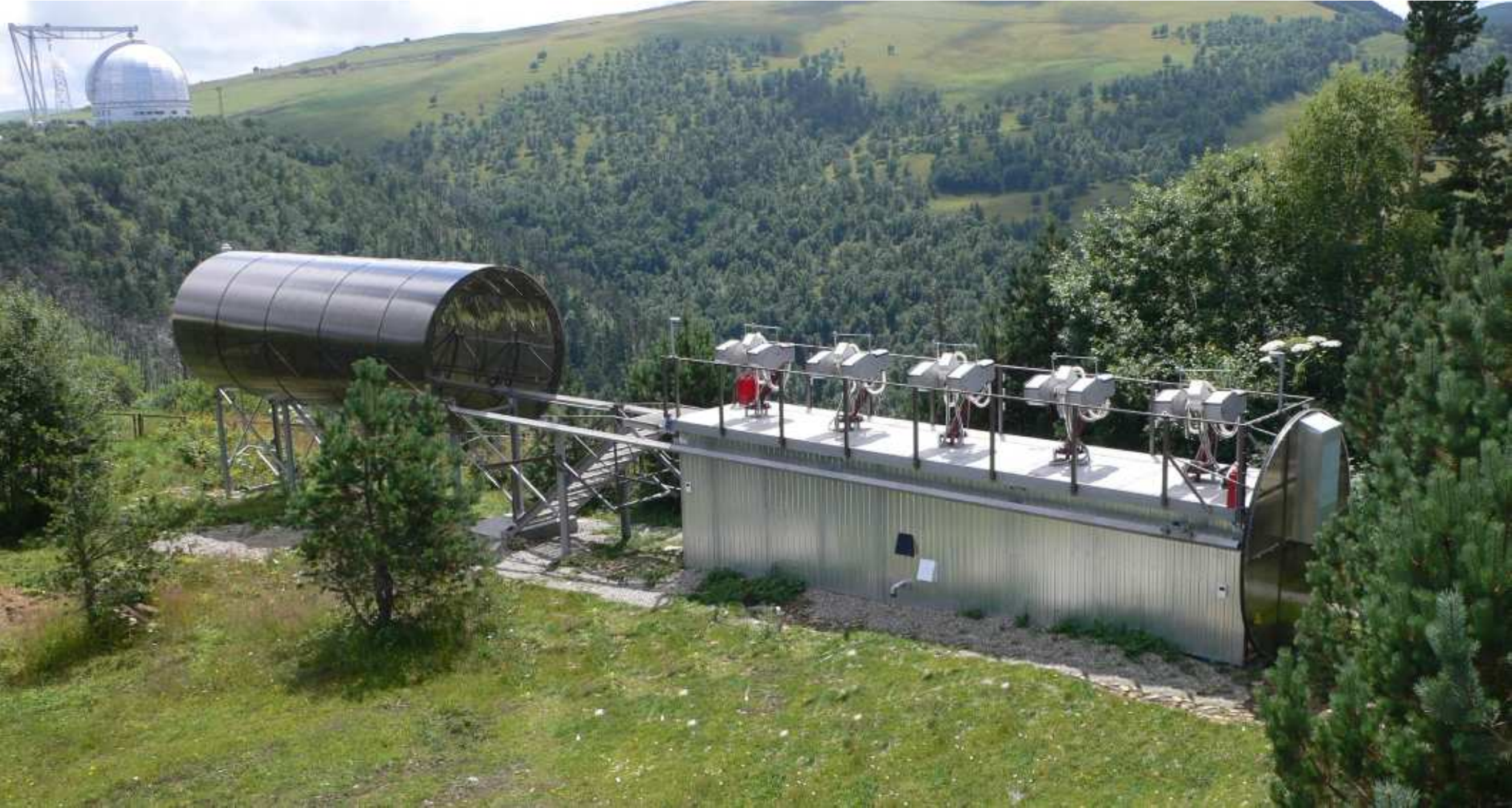}}
    \end{center}
    }
\vskip2mm
\captionb{3} {All the nine MMT channels installed on five mounts in a
single cylindrical dome, shown here open.  The Russian 6 m telescope is
seen in the background.} }
\end{figure}
\vskip2mm

The channels are equipped with Andor Neo sCMOS detectors having
$2560\times2160$ pixels, 6.4~$\mu$m each. The field of view of a
channel is roughly $9^\circ\times11^\circ$, with an angular
resolution of roughly $16''$ per pixel. The detector is able to
operate with exposure times as short as 0.03~s; in our work, we
use 0.1~s exposures providing us 10~frames per second.

Every channel is operated with a dedicated PC that controls its
hardware, acquires images from the detector and performs data
processing. The amount of data acquired by a single channel is about
3~Tb in eight hours of observations. The system as a whole is
controlled by a separate PC.

Initial tests show that the FWHM of stars as seen by MMT channels
is about 2~pixels wide. The detection limit with a signal-to-noise
ratio of 5 in white light for 0.1~s exposure time is close to
11~mag, when calibrating to $V$-band magnitudes, and to 12~mag in
the $B$~band.

\sectionb{3}{MMT FIRST LIGHT}

The MMT saw first light in March 2014 and since then has been in
test observations.

The MMT data processing software implements both a fast differential
imaging pipeline intended for detection of transient objects and a
slower image processing pipeline intended for performing astrometry
and photometry of the whole field of view in order to detect
variability on longer time scales.

   The first pipeline is fully implemented and is being extensively
tested, actively detecting meteor trails, satellite passes and rapid
flashes on the sky in the sub-second time domain.

\begin{figure}[!tH]
  \vbox{
    \begin{center}
\resizebox*{0.5\columnwidth}{!}{\includegraphics[angle=0]{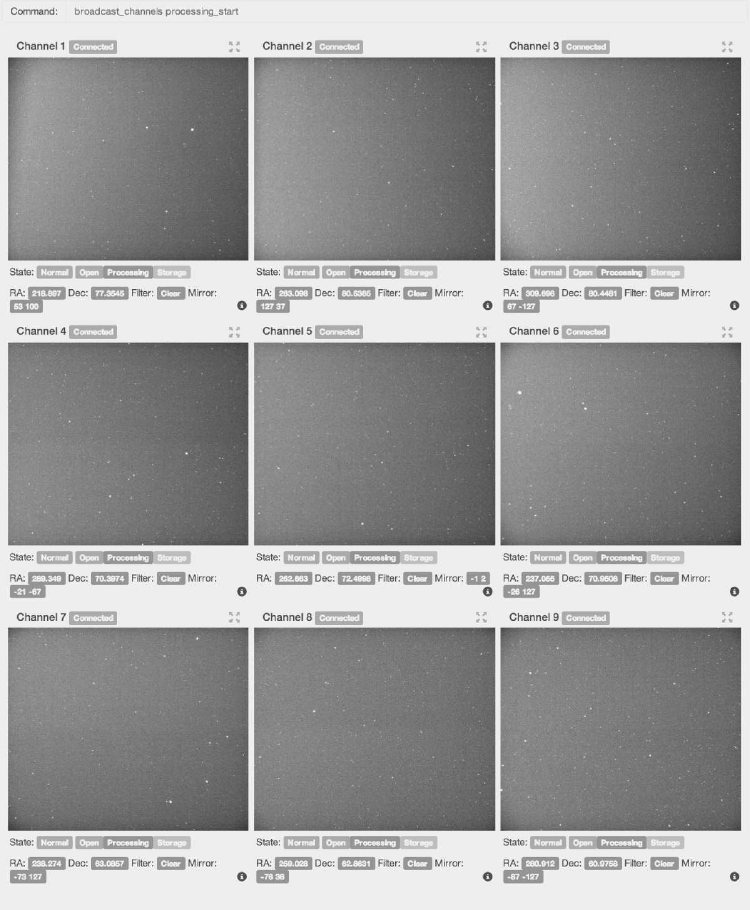}}
\vskip3mm
\resizebox*{0.5\columnwidth}{!}{\includegraphics[angle=0]{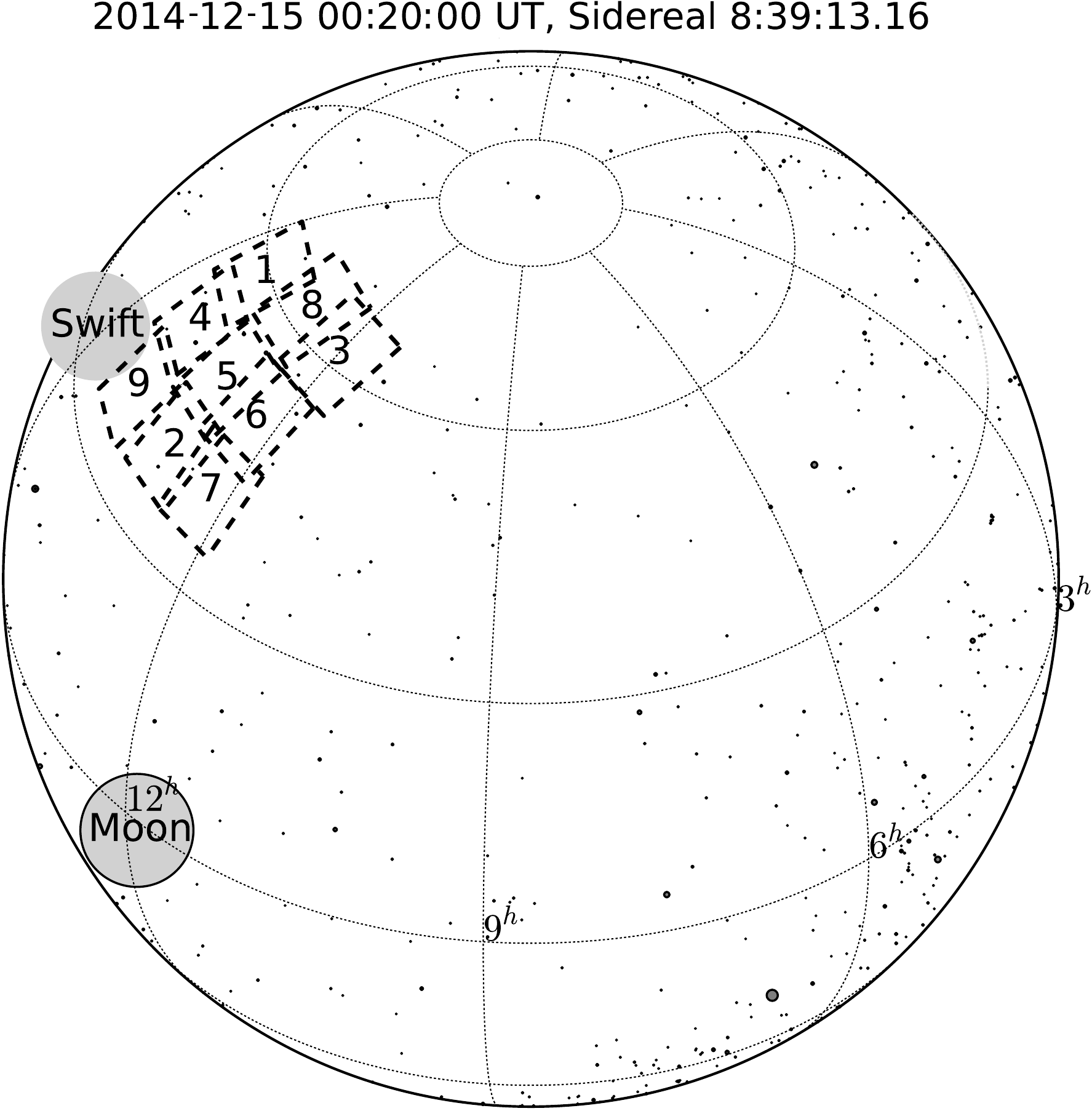}}
    \end{center}
\vskip2mm
\captionb{4} {Top panel:  the sky view of all nine MMT channels as
displayed by MMT control web interface.  Bottom:  actual placement of
nine fields of view on the sky in the monitoring mode. The actual
positions of the Moon and the Swift gamma-ray observatory viewfield are
marked by the gray circles.}
}
\end{figure}


   \begin{figure}[!tH]
  \vbox{
    \begin{center}
\resizebox*{\columnwidth}{!}{\includegraphics[angle=0]{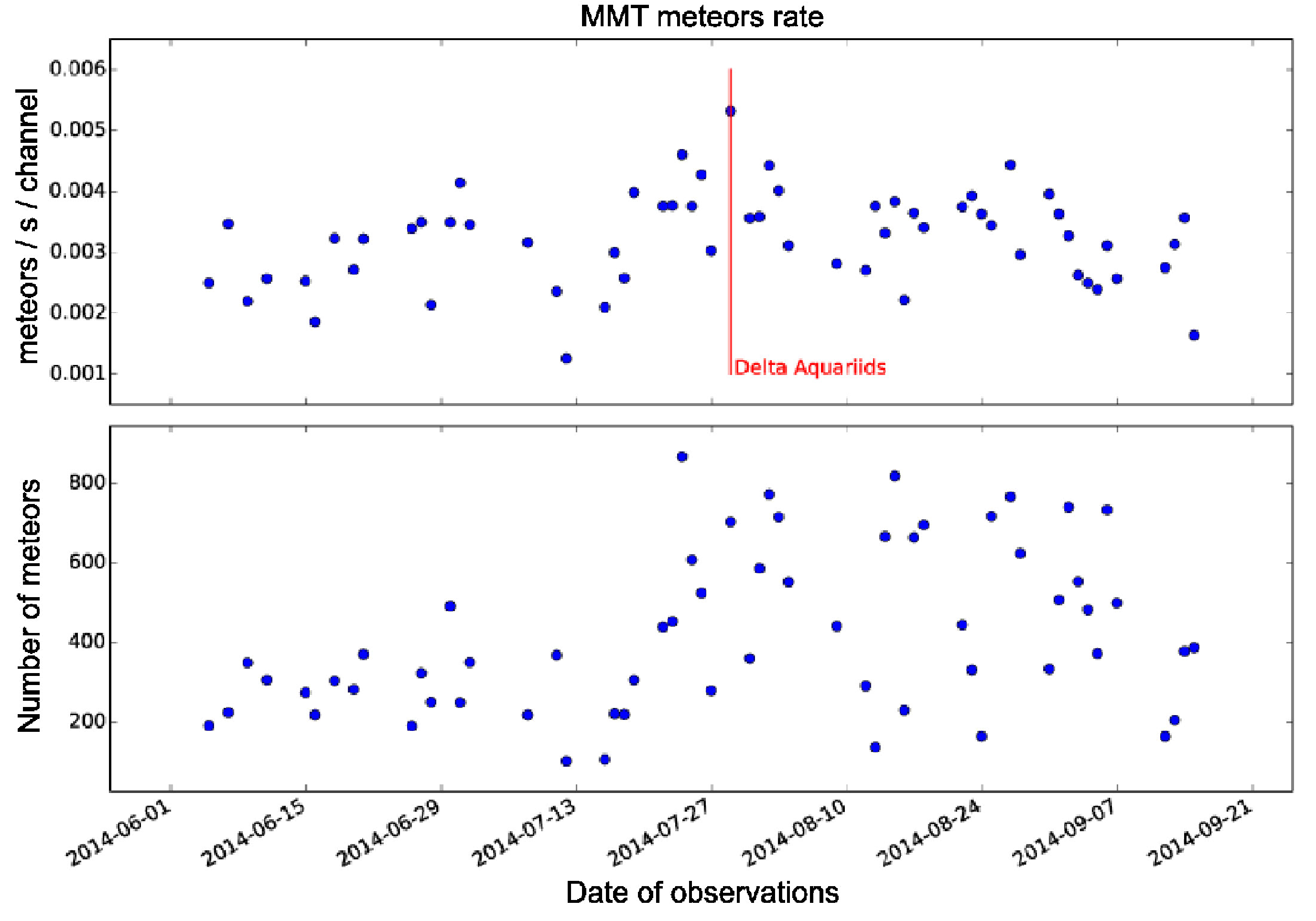}}
    \end{center}
    \vspace{-1mm}
\captionb{5} {Observations of meteors in monitoring mode:  rate of
events detected per second per channel (top) and total number of meteors
detected per night (bottom).  The rate of events is quite stable, while
the total number reflects both the duration of dark night time and
differences in weather conditions.  Small-scale peaks in the rate might
also correspond to small meteor streams.  } }
\end{figure}

 The principle used 
is, while working on high frame rate of 10 frames per second, to
build an iteratively-updated comparison image of the current field
of view using a numerically efficient running median algorithm, as
well as a threshold image using a similarly constructed running
\textit{median absolute deviation} estimate, and to compare each
frame to them, extracting candidate transient objects and analyzing
these objects from consecutive frames, as described in our previous
publications (see, e.g., Karpov et al. 2005).

The second, photometric pipeline is still in preparation, with
astrometric module already implemented (using Astrometry.net code
described in Lang et al. 2010) and providing astrometric solutions
for all the data acquired by MMT. The object detection and
photometry is, however, still in development, which is complicated
due to the relatively wide PSF of the Canon objective being used.

Below we briefly review various kinds of data being acquired.

\subsectionb{3.1}{Meteors}

Meteors are probably the most frequent astrophysical phenomena
flashing in the sky, and easiest to detect in the MMT data flow.
Detection of meteor trails is performed on a differential image
based on their typically elongated shapes. Then the elongated trails
from consecutive frames, having similar directions of elongation,
are being merged into single event. Dedicated analysis subroutine
then extracts the meteor trail using Hough transformation, detects
its range on every frame, and estimates the direction of meteor
motion and its velocity. The software also performs the search for
possible radiants of meteor streams by constructing a map of
pair-wise intersections of all meteor trails detected over the night
and studying its spatial distribution. Examples of such maps are
presented in Fig.~6.


\begin{figure}[!tH]
  \vbox{
    \begin{center}
\resizebox*{0.97\columnwidth}{!}{\includegraphics[angle=0]{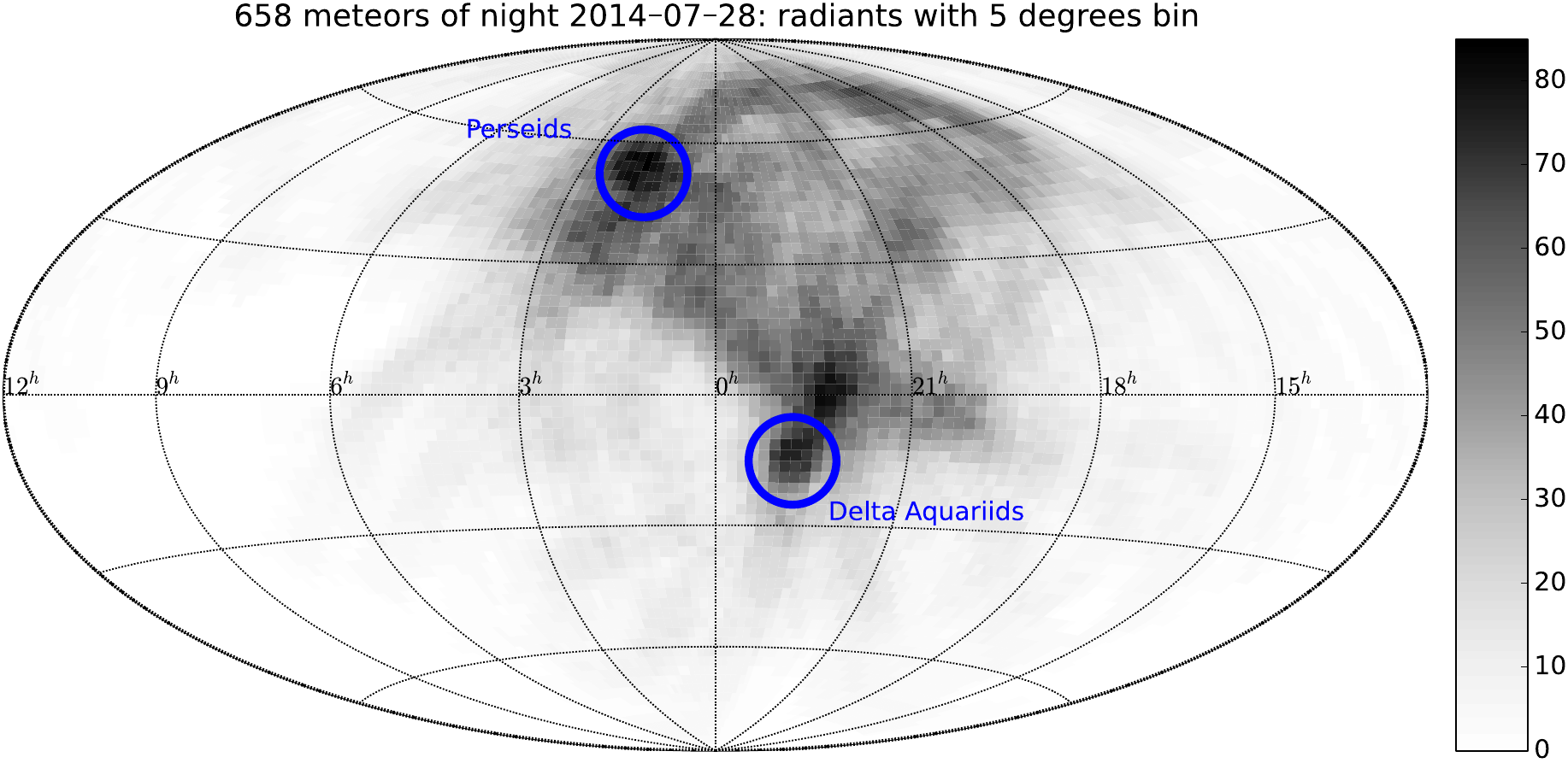}}
\vskip2mm
\resizebox*{0.99\columnwidth}{!}{\includegraphics[angle=0]{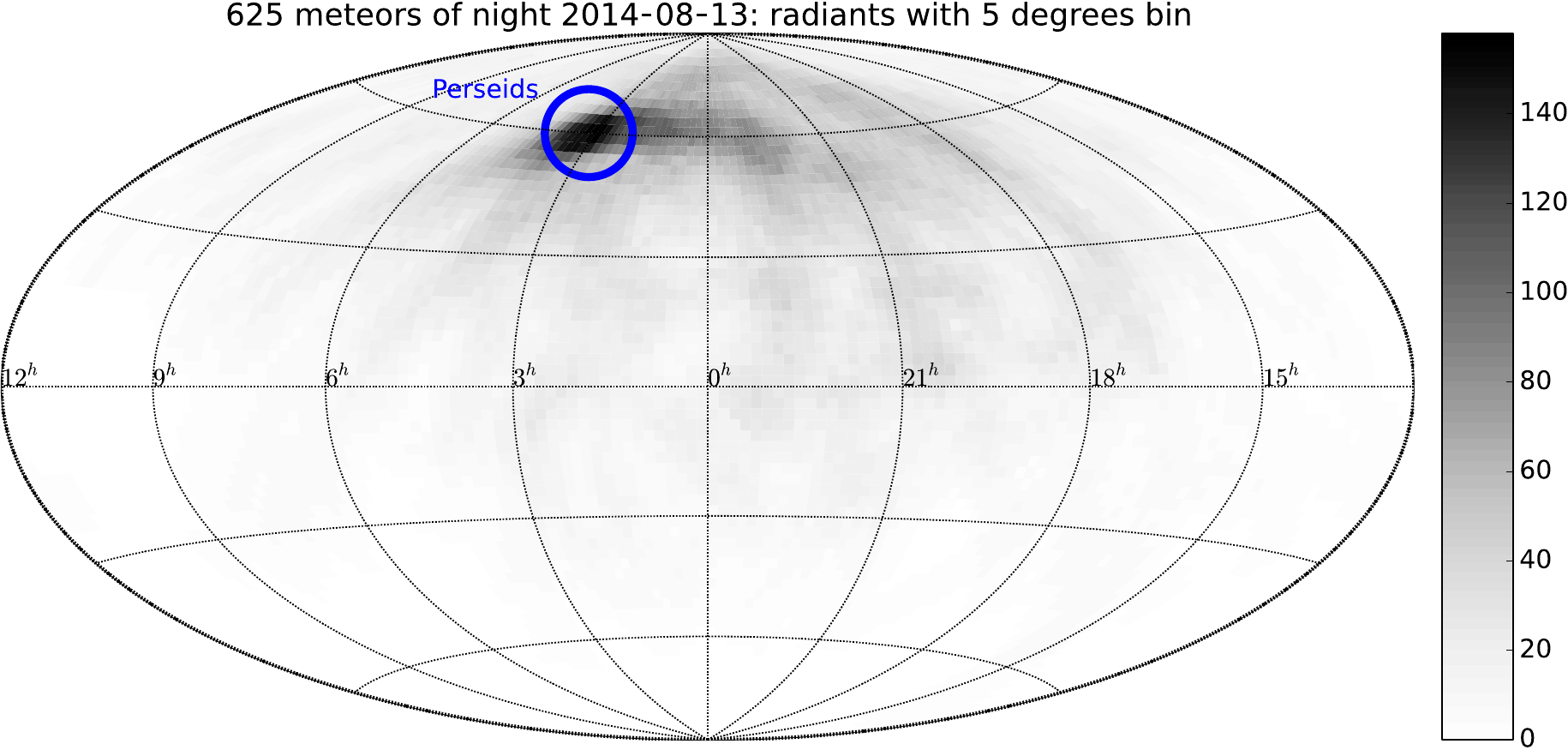}}
    \end{center}
    \vspace{-3mm}
\captionb{6} {Observations of meteors in monitoring mode.  Top:
pair-wise intersections (map of possible radiants) of meteor trails
detected on 2014 July~28 (the highest peak in the top panel of Fig.~5).
The radiant corresponding to the Delta~Aquariids meteor stream is
clearly seen, along with the onset of the Perseids meteor stream that
peaked two weeks later.  Bottom:  map of possible radiants of meteor
streams detected on 2014 August~13.  The radiant corresponding to the
Perseids meteor stream is evident.  } }
\end{figure}

The typical rate of meteors detected by the MMT is $\sim0.003$
events/s/channel (see Fig.~5), corresponding to $\sim700$ meteors
detected during an eight-hour dark night.

\subsectionb{3.1}{Satellites}

Detection of rapidly moving objects is implemented by comparing
lists of objects detected on consecutive differential frames and
extracting those that move along (nearly) straight lines with a
(slowly varying or) constant velocity. This is being done
iteratively, starting from the third appearance of the object on the
frame. After initial detection, the object is being tracked until it
fades below the detection limit or leaves the field of view;
afterwards, its  trajectory and  light curve are stored in the
database.


\begin{figure}[!tH]
  \vbox{
    \begin{center}
\resizebox*{0.98\columnwidth}{!}{\includegraphics[angle=0]{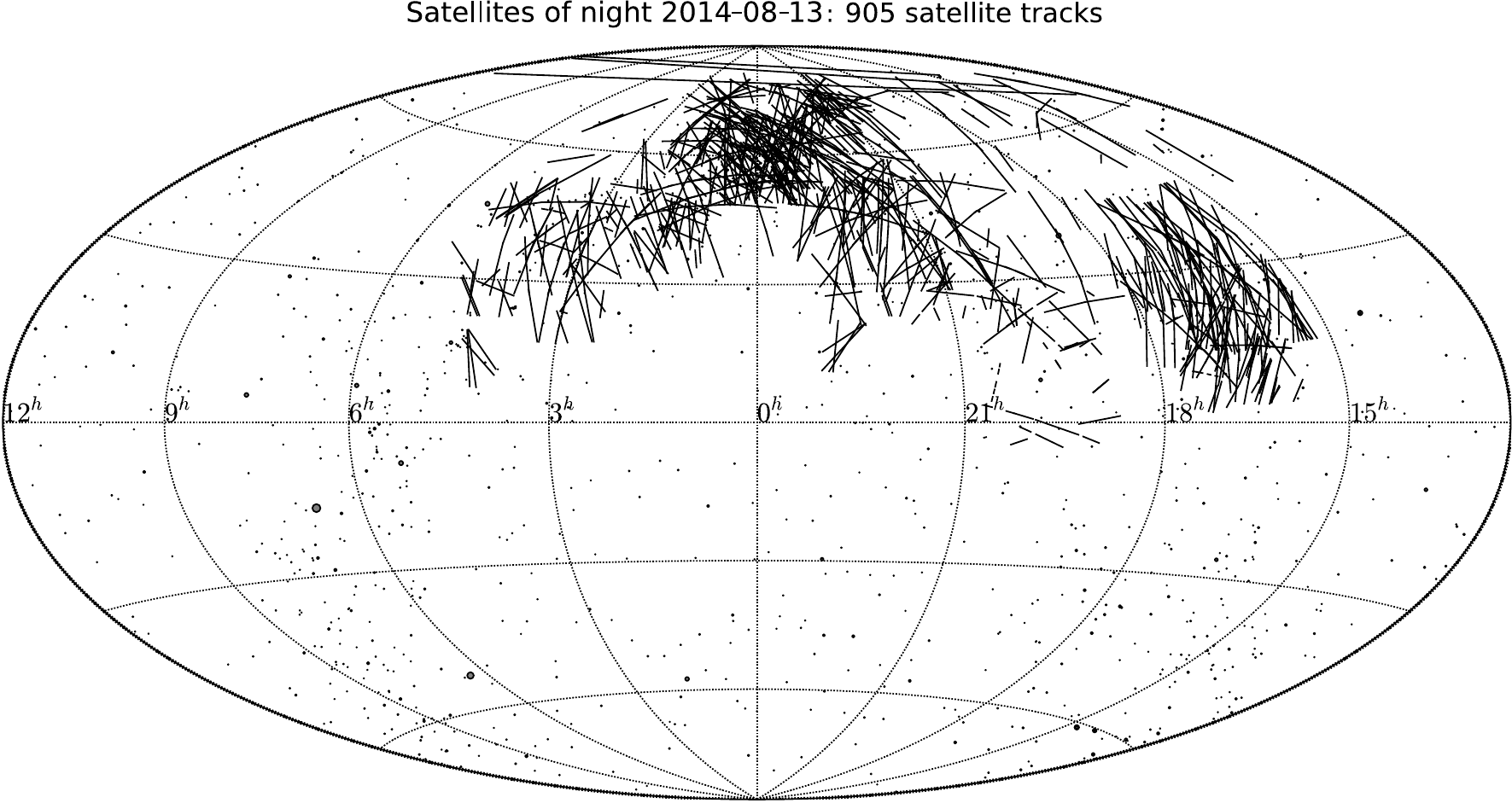}}
    \end{center}
    \vspace{-4mm}
\captionb{7} {Satellite trails detected by the MMT on the night of 2014
August 13.  Only trajectories with more than 100 detection points are
displayed.} }
\end{figure}


\begin{figure}[!tH]
\vbox{\vspace{2mm}
\centerline{\psfig{figure=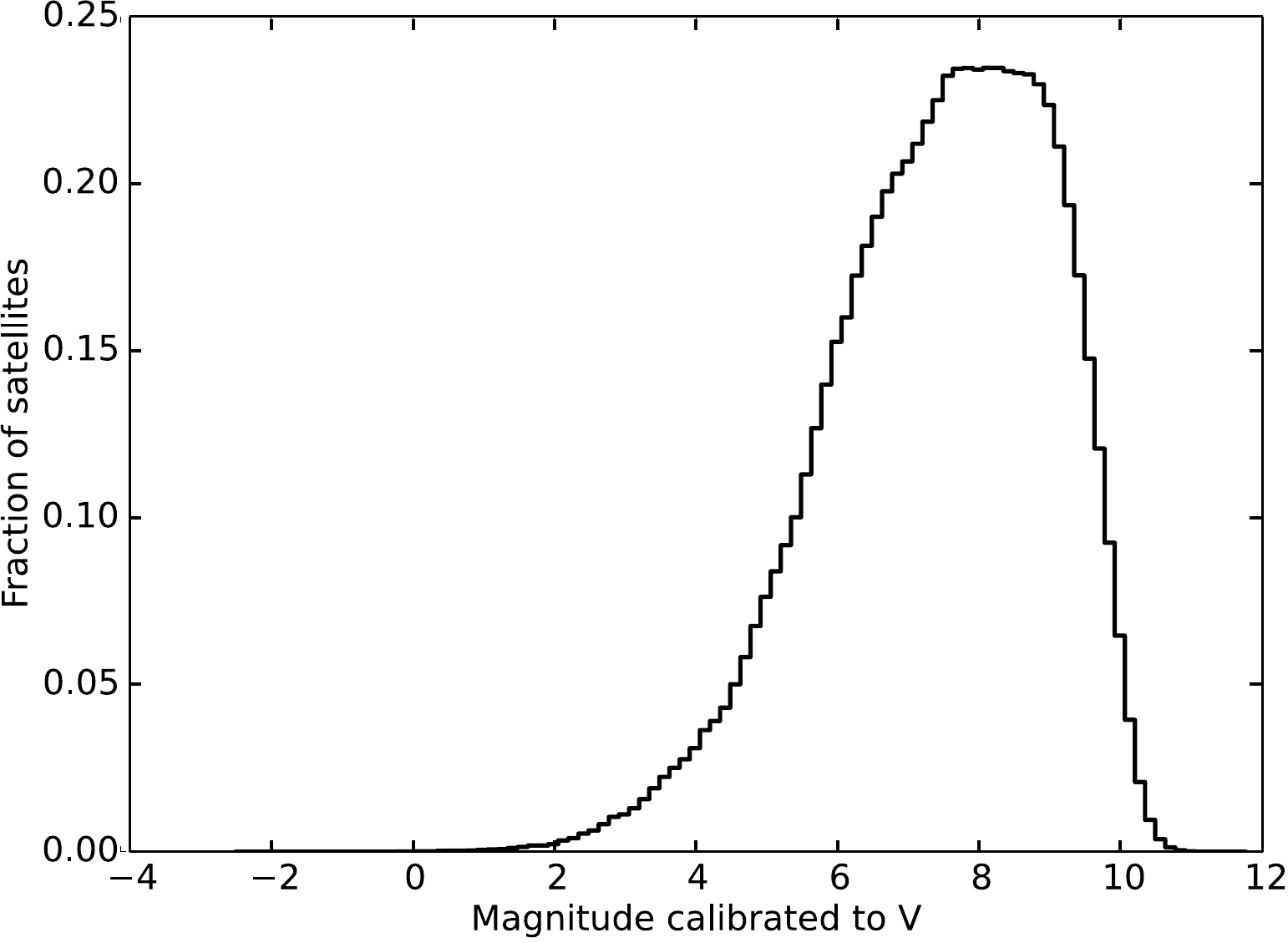,width=80mm,angle=0,clip=}}
\vspace{2mm}
\captionb{8}{Overall distribution of mean magnitudes of
satellite tracks detected over seven months of MMT operation.}}
\end{figure}



\begin{figure}[!tH]
  \vbox{
\vskip2mm
    \begin{center}
\resizebox*{\columnwidth}{!}{\includegraphics[angle=270]{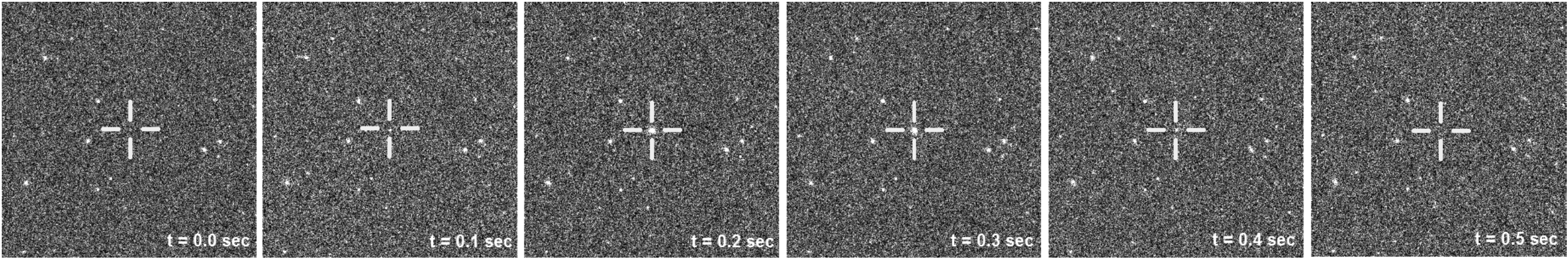}}
    \end{center}
    \vspace{-1mm}
    \captionb{9}
{A rapid optical flash detected by the MMT, with a duration less than
0.5~s and peak brightness reaching $\sim$\,6.5 mag.  The flash coincides
with a high-altitude passage of MOLNIYA satellite, according to the
NORAD database (predicted distance is less than $400''$, a typical error
for satellite positions).} }
\end{figure}

An example of satellite trails detected by the MMT on a single night
is displayed in Fig.~7, and the distribution of their mean
magnitudes over seven months of MMT operation is shown in Fig.~8.


\subsectionb{3.3}{Fast optical bursts}

The original aim of the MMT differential imaging pipeline is
detection of rapid optical flashes of astrophysical origin, which is
being performed by detecting star-like objects visible on several
consecutive differential images (to filter out sporadic noise events
and cosmic rays) and not changing their position. As of now, we are
still in process of calibrating this part of the pipeline, as it is
highly contaminated by stellar scintillations and detector noise
spikes. We are, however, able to detect a number of rapid flashes
caused by rotation of high-altitude, slowly moving satellites, which
produce short (to half a second) events with negligible motion. Such
flashes are practically indistinguishable from anticipated
astrophysical bursts and can be filtered out only by comparing their
positions to those predicted for known satellites, what is being
done in the MMT software using the NORAD database.

An example of such an event is shown in Fig.~9.

As of now, we did not detect any rapid flashes not coincident with
a high-altitude satellite and not having a light curve identical
to those produced by such satellites.

\sectionb{4}{CONCLUSIONS}

The Mini-MegaTORTORA (MMT) instrument is already operational and
shows the performance close to that expected. We hope it will be
useful for studying various phenomena on the sky, both astrophysical
and artificial in origin. We expect it to be used for studying faint
meteoric streams crossing the Earth's orbit, for detecting new
comets and asteroids, as well as for finding flashes of flaring
stars and novae, studying variable stars of various types, detecting
transits of exoplanets, searching for bright supernovae and optical
counterparts of gamma-ray bursts. It will also be used for citizen
science by providing imaging capabilities for users of the GLORIA
robotic telescopes network project.

The novelty of the MMT is its ability to re-configure itself from
a wide-field to narrower-field instrument, which may open new ways
of studying the sky, as it may, in principle, autonomously perform
a thorough study of objects it discovers -- simultaneously acquire
three-color photometry and polarimetry of them. We will
demonstrate MMT performance in such a mode in subsequent papers.


\thanks{The construction of the Mini-MegaTORTORA is performed according
to the Russian Government Program of Competitive Growth of Kazan Federal
University and partially supported by the grants RFBR 09-02-12053 and
12-02-00743.  International contacts within the project were supported
by the European Union Seventh Framework Programme (FP7/2007-2013) under
the grant agreement 283783 (GLORIA project).  The theoretical analysis
for the basic observational programs of MMT was performed under the
financial support of the grant of Russian Science Foundation No.
14-50-00043.  }

\References

\refb Beskin G. M., Karpov S. V., Bondar S. F. et al. 2010a, Phys.
Usp., 53, 406

\refb Beskin G., Karpov S., Bondar S. et al. 2010b, ApJ, 719, L10

\refb Karpov S., Beskin G., Biryukov A. et al. 2005, Nuovo Cimento
C, 28, 747

\refb Karpov S., Beskin G., Bondar S. et al. 2008, GRB Coordinates
Network Circular, 7452, 1

\refb Karpov S., Beskin G., Bondar S. et al., 2012, Astronomical
Society of India Conf. Ser., 7, 219

\refb Lang D., Hogg D. W., Mierle K. et al. 2010, AJ, 139, 1782

\refb Molinari E., Bondar S., Karpov S. et al. 2006, Nuovo Cimento
B, 121(12), 1525

\refb Racusin J. L., Karpov S. V., Sokolowski M. et al. 2008,
Nature, 455, 183

\refb Zolotukhin I., Beskin G., Biryukov A. et al. 2004, AN, 325,
675

\end{document}